\documentclass[aps,twocolumn,groupedaddress]{revtex4}
\usepackage{amsmath,amssymb,epsfig,natbib,bm}

\def\be{\begin{equation}}
\def\ee{\end{equation}}
\def\bea{\begin{eqnarray}}
\def\eea{\end{eqnarray}}
\def\biposh#1#2#3{ A^{#1}_{#2|#3}}

\begin{document}

\newcommand{\edth}{\,\eth\,}
\renewcommand{\beth}{\,\overline{\eth}\,}

\title{Testing Global Isotropy of Three-Year Wilkinson Microwave Anisotropy Probe (WMAP) Data: Temperature Analysis}
\author{Amir Hajian}
\affiliation{ Department of Physics, Jadwin Hall, Princeton University,   PO Box 708, Princeton, NJ 08542.  \\
Department of Astrophysical Sciences, Peyton Hall, Princeton University,
Princeton, NJ 08544. }
\author{Tarun Souradeep}
\affiliation{Inter-University Centre for Astronomy and Astrophysics
(IUCAA), Ganeshkhind, Pune-411007, India.}
\date{July 7 2006}
\begin{abstract}
We examine statistical isotropy of large scale anisotropies of the
Internal Linear Combination (ILC) map, based on three year WMAP
data. Our study reveals no significant deviation from statistical
isotropy on large angular scales of 3-year ILC map. Comparing
statistical isotropy of 3-year ILC map and 1-year ILC map, we find a
significant improvement in 3-year ILC map which can be due to the
gain model, improved ILC map processing and foreground minimization.
\end{abstract}

\maketitle

\section{Introduction}
The Cosmic Microwave Background (CMB) anisotropy has been shown to
be a very powerful observational probe of cosmology. Detailed
measurements of the anisotropies in the CMB can provide a wealth of
information about the global properties, constituents and history
 of the Universe. In standard cosmology, the CMB
anisotropy is expected to be statistically isotropic, {\it i.e.},
statistical expectation values of the temperature fluctuations  (and
in particular the angular correlation function) are preserved under
rotations of the sky. This property of  CMB anisotropy has been
under scrutiny after the release of the first year of WMAP data
\cite{erik04a, Copi:2003kt, Schwarz:2004gk, Hansen:2004vq, angelwmap, Land:2004bs,Land:2005ad, Land:2005dq, Land:2005jq,Land:2005cg, Bielewicz:2004en, Bielewicz:2005zu,Copi:2005ff, Copi:2006tu, Naselsky:2004gm, Prunet:2004zy,Gluck:2005td, Stannard:2004yp, Bernui:2005pz, Bernui:2006ft, SadeghMovahed:2006em, Freeman:2005nx, us_apj, Chen:2005ev}. We use a method based on bipolar expansion of
the two point correlation function which is an improved and enhanced
follow up on our previous work on first year WMAP data \cite{us_apj,
us_bigpaper}. This method is  shown to be sensitive to structures
and patterns in the underlying two-point correlation function.

We apply our method to the improved Internal Linear Combination
(ILC) map \cite{Hinshaw:2006ia}, based on three year WMAP
data\footnote{This map is available on LAMBDA  as part of the
three-year data release \cite{Hinshaw:2006ia,Jarosik:2006ib,Page:2006hz,Spergel:2006hy}.}. We choose the ILC map for testing
statistical isotropy (SI) of the CMB anisotropy for the following
reasons
  \begin{enumerate}
  \item{The ILC is a full-sky map and hence is easier to work with. Masking the sky results in violation of statistical isotropy. An originally SI CMB anisotropy map deviates from SI after masking \cite{us_bigpaper}.  }
   \item{Residuals from Galactic removal errors in the three-year ILC map are estimated to be less than 5 $\mu K$ on angular scales greater than $\sim10\deg$ \cite{Hinshaw:2006ia}.
   Hence at low-$l$, multipoles are not significantly affected by foregrounds.
   In addition, it is interesting to examine the above statement by testing the statistical isotropy of the ILC map.}
  \item{On large scales, the three-year ILC map is believed to provide a reliable estimate of the CMB signal, with negligible instrument noise, over the full sky \cite{Hinshaw:2006ia}. }
  \end{enumerate}
  These properties of the ILC map allow us to study the cosmological signal on large scales.
In addition, there are theoretical motivations for hunting for SI violation on large scales of CMB anisotropy. Topologically compact spaces~\cite{ell71, Lachieze-Rey:1995kj, lev02, linde} and anisotropic cosmological models~\cite{ellis&maccallum, collins&hawking1, collins&hawking2, doroshke, barrow, Jaffe:2006fh, us_bianchi} are examples of this. Both observational artifacts and the above theoretical models cause a departure from statistical isotropy and it has been shown that our method is a useful tool to find out these deviations (see {\it e.g.} \cite{us_bianchi, us_foregrounds, us_prl}). The rest of this paper is organized as follows:  Section \ref{characterization}  is a brief introduction to temperature anisotropy of CMB. Section \ref{SIsection} describes the formulation of statistical isotropy in general. Section \ref{estimators} is a description of estimators we use to look for deviations from statistical isotropy. We present the application of our method on the WMAP data in Section \ref{data}. Section \ref{discussion} contains discussion on the cosmological implications of our null detection of deviations from statistical isotropy on large angular scales in 3-year ILC map of WMAP data, and in Section \ref{summary} we summarize our results.

\section{Characterization of CMB temperature anisotropy}
\label{characterization} The CMB anisotropy is fully described by
its temperature anisotropy and polarization. The temperature
anisotropy is a scalar random field, $\Delta
T(\hat{n})=T(\hat{n})-T_0$, on a 2-dimensional surface of a sphere
(the sky), where $\hat{n}=(\theta,\phi)$ is a unit vector on the
sphere and $T_0=\int{\frac{d\Omega_{\hat{n}}}{4\pi}T(\hat{n})}$
represents the mean temperature of the CMB. It is convenient to
expand the temperature anisotropy field into spherical harmonics,
the orthonormal basis on the sphere, as \be \label{yelemexpand}
\Delta T(\hat{n}) \, = \, \sum_{l,m} a_{lm}Y_{lm}(\hat{n}) \,\,, \ee
where the complex quantities, $a_{lm}$ are given by \be \label{alm}
a_{lm} =\int{\mathrm{d} \Omega_{\hat{n}}Y_{lm}^{*}(\hat{n}) \Delta
T(\hat{n})}. \ee Statistical properties of this field can be
characterized by $n$-point correlation functions \be \label{npoint}
\langle \Delta T(\hat{n}_1) \Delta T(\hat{n}_2)\cdots \Delta
T(\hat{n}_n)\rangle. \ee Here the bracket denotes the ensemble
average, {\it{i.e.}} an average over all possible configurations of
the field. CMB anisotropy is believed to be Gaussian
\cite{Bartolo:2004,Komatsu:2003}.
 Hence the connected part of $n$-point functions
disappears for $n > 2$. Non-zero (even-$n$)-point correlation
functions can be expressed in terms of the $2$-point correlation
function. As a result, a Gaussian distribution is completely
described by the two-point correlation function \be
C(\hat{n},\hat{n'})\,=\, \langle \Delta T(\hat{n}) \Delta
T(\hat{n}') \rangle. \ee Equivalently; as it is seen from
eqn.~(\ref{alm}), for a Gaussian CMB anisotropy, $a_{lm}$ are
complex Gaussian random variables too. Therefore, the {\it
covariance matrix}, $\langle a_{lm}a^{*}_{l^\prime
m^\prime}\rangle$, fully describes the whole field. Throughout this
paper we assume Gaussianity to be valid.
\section{Statistical isotropy}
\label{SIsection} Two point correlations of CMB anisotropy,
$C(\hat{n}_1,\, \hat{n}_2)$, are two point functions on $S^2 \times
S^2$, and hence can be expanded as \be \label{bipolar}
C(\hat{n}_1,\, \hat{n}_2)\, =\, \sum_{l_1,l_2,\ell,M} \biposh{}{\ell
M}{l_1 l_2} Y^{l_1l_2}_{\ell M}(\hat{n}_1,\, \hat{n}_2). \ee Here
$\biposh{}{\ell M}{l_1 l_2}$ are coefficients of the expansion
(hereafter BipoSH coefficients) and $Y^{l_1l_2}_{\ell
M}(\hat{n}_1,\, \hat{n}_2)$ are bipolar spherical harmonics defined
by eqn.~(\ref{bipolars}). Bipolar spherical harmonics form an
orthonormal basis on $S^2 \times S^2$ and transform in the same
manner as the spherical harmonic function with $\ell,\, M$ with
respect to rotations \cite{Var}. One can inverse-transform
$C(\hat{n}_1,\, \hat{n}_2)$ in eqn.~(\ref{bipolar}) to get the
coefficients of expansion, $\biposh{}{\ell M}{l_1 l_2}$, by
multiplying both sides of eqn.(\ref{bipolar}) by
$Y^{*l'_1l'_2}_{\ell'M'}(\hat{n}_1,\hat{n}_2)$ and integrating over
all angles. Then the orthonormality of bipolar harmonics, eqn.
(\ref{A2}), implies that \be \label{alml1l2} \biposh{}{\ell M}{l_1
l_2} \,=\,\int d\Omega_{\hat{n}_1}\int d\Omega_{\hat{n}_2} \,
C(\hat{n}_1,\, \hat{n}_2)\, Y^{*l_1l_2}_{\ell
M}(\hat{n}_1,\hat{n}_2). \ee The above expression and the fact that
$C(\hat{n}_1,\, \hat{n}_2)$ is symmetric under the exchange of
$\hat{n}_1$ and $\hat{n}_2$ leads to the following symmetries of
$\biposh{}{\ell M}{l_1 l_2}$ \bea \label{sym} \biposh{}{\ell M}{l_2
l_1} \,&=&\,(-1)^{(l_1+l_2-L)}\biposh{}{\ell M}{l_1 l_2}, \\
\nonumber \biposh{}{\ell M}{ll} \, &=& \, \biposh{}{\ell M}{ll}
\,\,\delta_{\ell,2k}, \,\,\,\,\,\,\,\,\,\,\,\,\,\,\,\,\,\,\,\,\,
k=0,\,1,\,2,\,3,\,\cdots.  \eea It has been shown \cite{us_bigpaper}
that Bipolar Spherical Harmonic (BipoSH) coefficients,
$\biposh{}{\ell M}{l_1 l_2}$, are in fact linear combinations of
off-diagonal elements of the covariance matrix, \be \label{ALMvsalm}
 \biposh{}{\ell M}{l_1 l_2} \,=\, \sum_{m_1m_2}
\langle a_{l_1m_1}a^{*}_{l_2 m_2}\rangle (-1)^{m_2} {\mathcal
  C}^{\ell M}_{l_1m_1l_2 -m_2}.
  \ee
  where ${\mathcal  C}_{l_1m_1l_2m_2}^{\ell M}$ are Clebsch-Gordan coefficients (see the Appendix).This
  clearly shows that $\biposh{}{\ell M}{l_1 l_2}$ completely  represent the information of the covariance matrix.
  When statistical  isotropy holds, it is guaranteed that the covariance matrix is  diagonal,
  \be
  \langle a_{lm}a^{*}_{l' m'}\rangle = C_{l}\,\,
  \delta_{ll^\prime} \delta_{mm'}
  \ee
  and hence the angular power
  spectra carry all information of the field. Substituting this into
  eqn.~(\ref{ALMvsalm}) gives
   \be \label{SIALM1}
   \biposh{}{\ell M}{ll'} \,=\,(-1)^l C_{l} (2l+1)^{1/2} \, \delta_{ll^\prime}\, \delta_{\ell 0}\, \delta_{M0}.
   \ee
   The above expression tells us
  that when statistical isotropy holds, all BipoSH coefficients,
  $\biposh{}{\ell M}{ll'}$, are zero except those with $\ell=0,
  M=0$ which are equal to the angular power spectra up to a $(-1)^l
  (2l+1)^{1/2}$ factor. BipoSH expansion is the most general way of
  studying two point correlation functions of CMB anisotropy. The well   known angular power spectrum, $C_l$ is in fact a subset of the   corresponding BipoSH coefficients,
  \be \label{SIALM}
  C_{l}\,=\, \frac{(-1)^{l}}{\sqrt{2l+1}} \biposh{}{0 0}{ll}.
  \ee
  Therefore to test a CMB map for statistical isotropy, one should  compute the BipoSH coefficients for the maps and look for   nonzero BipoSH coefficients. Statistically significant deviations from zero would mean violation of   statistical isotropy.
   \section{Estimators}
\label{estimators}
 Given a CMB anisotropy map, one can measure BipoSH coefficients bye the following estimator\footnote{In statistics, an estimator is a function of the known
data that is used to estimate an unknown parameter; an estimate is
the result from the actual application of the function to a
particular set of data. Many different estimators may be possible
for any given parameter.} \be \label{estimator} \biposh{}{\ell
M}{ll'} =\sum_{m m^\prime} \sqrt{W_l W_{l'}}\, a_{lm}a_{l^\prime
m^\prime} \, \, {\mathcal{ C}}^{\ell M}_{lml^\prime m^\prime}\,\quad
, \ee
  where $W_l$ is the Legendre
transform of the window function. The above estimator is a
combination of $C_l$ and hence is un-biased\footnote{By bias we mean
the mismatch between ensemble average of the estimator and the true
value.}. However it is impossible to measure all $\biposh{}{\ell
M}{ll'}$ individually because of cosmic variance. Combining BipoSH
coefficients helps to reduce the cosmic variance\footnote{This is
similar to combining $a_{lm}$ to construct the angular power
spectrum, $C_l=\frac{1}{2l+1}\sum_{m}{|a_{lm}|^2}$, to reduce the
cosmic variance.}. There are several ways of combining BipoSH
coefficients. Here we choose two methods.
\begin{figure}[t]
\includegraphics[scale=0.3  , angle=0]{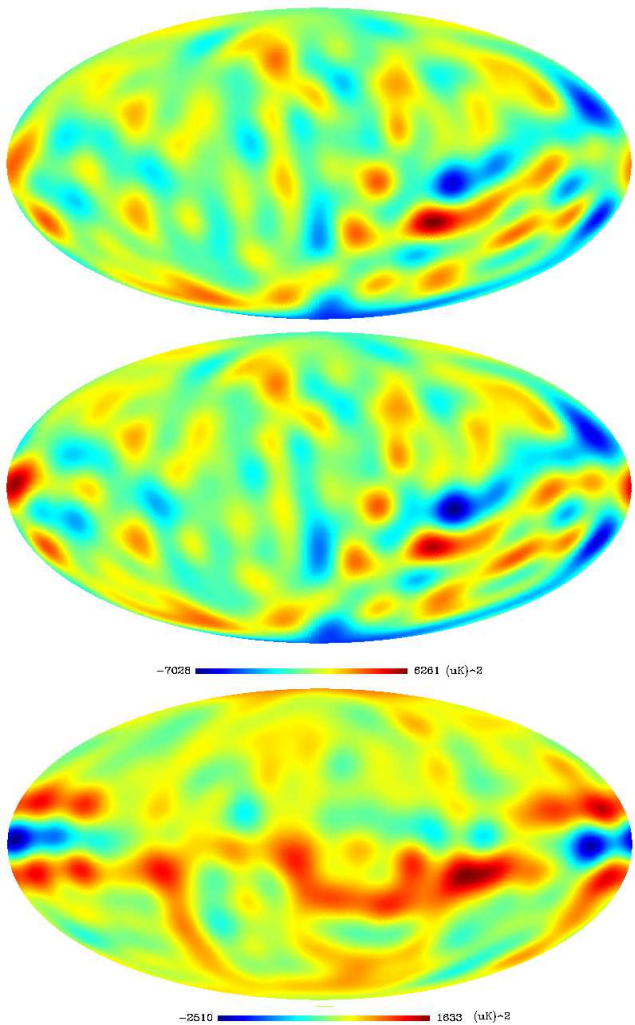}
    \caption{{\bf Top:} A {\it bipolar} map generated from bipolar coefficients, $A_{\ell M}$, of 3-year ILC map.
    {\bf Middle:}  bipolar map based on 1-year ILC map and {\bf Bottom:} differences between the two maps (note the scales).
    The top map (ILC-3) has smaller fluctuations comparing to the middle one (ILC-1) except for the hot spot near the equator.
    Differences between these two maps mostly arise from a band around the equator {\it in bipolar space}.
    Both ILC maps are smoothed by a band pass filter, $W^{S}(l_t=2, l_s=10)$. }
  \label{bipolarmap}
\end{figure}
\subsection{First Method: BiPS}
 Among the several possible combinations of BipoSH
coefficients, the Bipolar Power Spectrum (BiPS) was proved to be a
useful tool with interesting features \cite{us_apjl}. BiPS of CMB
anisotropy is defined as a convenient contraction of the BipoSH
coefficients \be \label{kappal} \kappa_\ell \,=\, \sum_{l,l',M}
\left|\biposh{}{\ell M}{ll'}\right|^2 \geq 0. \ee BiPS is
interesting because it is orientation independent, {\it i.e.}
invariant under rotations of the sky. For models in which
statistical isotropy is valid, BipoSH coefficients are given by
eqn~(\ref{SIALM}), and therefore SI condition implies a null BiPS,
{\it i.e.}  $\kappa_\ell\,=\,0$ for every $\ell>0$, \be
\kappa_\ell\,=\,\kappa_0 \delta_{\ell 0}. \ee Non-zero components of
BiPS imply break down of statistical isotropy, and this introduces
BiPS as a measure of statistical isotropy. It is worth noting that
although BiPS is quartic in $a_{lm}$, it is designed to detect SI
violation and not non-Gaussianity \cite{us_bigpaper, us_apjl,
us_pramana, us_jgrg, us_apj}. An un-biased estimator of BiPS is
given by \be \label{estimatork} \tilde\kappa_\ell = \sum_{ll^\prime
M} \left|\biposh{}{\ell M}{ll'}\right|^2 - {\mathfrak B}_\ell\, ,
\ee where ${\mathfrak B}_\ell$ is the bias that arises from the SI
part of the map and is given by the angular power spectrum, $C_l$,
\bea \label{klisobias} {\mathfrak B}_\ell & \equiv &
\langle\tilde\kappa_\ell^B\rangle_{_{\rm SI}} \\ \nonumber & = &
(2\ell+1)\,\sum_{l_1} \sum_{l_2=|\ell-l_1|}^{\ell+l_1} W_{l_1}
W_{l_2} \times
\\ \nonumber & & \,\,\,\,\,\,\,\,\,\,
\,\,\,\,\,\,\,\,\,\,\left[ C_{l_1} C_{l_2} + (-1)^{\ell}\,
\delta_{l_1 l_2} \left(C_{l_1}\right)^{2} \right]\,. \eea The above
expression for ${\mathfrak B}_\ell$ is obtained by assuming
Gaussian statistics of the temperature fluctuations \cite{us_apjl,us_bigpaper}. 
\subsection{New Method: Reduced Bipolar Coefficients}
The BipoSH coefficients of eqn.~(\ref{estimator}) can be summed over
$l$ and $l'$ to reduce the cosmic variance, \be \label{first}
A_{\ell M}= \sum_{l=0}^{\infty}\sum_{l'=|\ell-l|}^{\ell+l}
\biposh{}{\ell M}{ll'}. \ee These reduced bipolar coefficients,
$A_{\ell M}$, by definition respect the following symmetry: \be
\label{almsymmetry} A_{\ell M}=(-1)^M A_{\ell -M}^*, \ee
 which indicates $A_{\ell 0}$ are always real. When SI condition
is valid, the ensemble average of $A_{\ell M}$ vanishes for all
$\ell$ and $M$ \be \langle A_{\ell M} \rangle = 0. \ee In any given
CMB anisotropy map, $A_{\ell M}$ would fluctuate about zero. A
severe breakdown of statistical isotropy will result in huge
deviations from zero. Reduced bipolar coefficients are not
rotationally invariant, hence they assign direction to the
correlation patterns of a map. We can combine $A_{\ell M}$ further
to define a power spectrum similar to how $a_{lm}$ are combined to
construct the angular power spectrum, $C_l$. We define \be
\label{dell} D_{\ell} = \frac{1}{2\ell +1}\sum_{M=-\ell}^{\ell}
A_{\ell M} A_{\ell M}^*. \ee The above estimator is rotationally
invariant. It has a positive bias and hence it has similar issues
that have been addressed for the BiPS studies earlier. This means
although the ensemble average of $A_{\ell M}$ for a statistically
isotropic case is zero, ensemble average of $D_{\ell}$ is always
greater than zero. However a major deviation from statistical
isotropy will result in a big $D_{\ell}$ (compared to that of a SI
case). In section \ref{data} we compare $A_{\ell M}$\ of ILC map
against an average of 1000 simulations of statistically isotropic
maps. We defer detailed studies of $D_\ell$ to the future
publication.

\begin{figure}[t]
\includegraphics[scale=0.3, angle=-90]{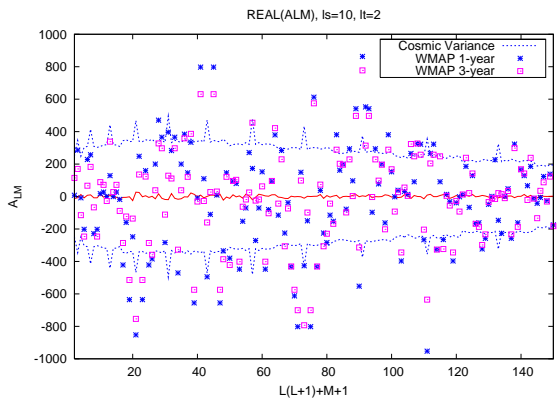}
    \caption{Real part of $A_{\ell M}$'s of ILC-3 (red square points) and ILC-1 (blue stars) for a $W^S(10,2)$
    filter that roughly keeps multipoles between 2 and 15.
     $\ell$ and $M$ indices are combined to a single index $n=\ell(\ell+1)+M+1$ and the blue dotted lines define
      1-$\sigma$ error bars derived from 1000 simulations of SI CMB anisotropy maps. Almost all of $A_{\ell M}$'s
       of ILC-3 are smaller than those of ILC-1 which means ILC-3 is more consistent with statistical isotropy.}
  \label{ALM_summed}
\end{figure}
\section{Application to the WMAP Data}
\label{data}
We carry out our analysis on 3-year ILC map and compare it to 1-year ILC map. In order to attribute a statistical significance to our results, we compare our results to 1000 simulations of SI CMB maps.  $a_{lm}$'s of these maps are generated up to an $l_{max}$ of 1024
(corresponding to HEALPix\footnote{http://healpix.jpl.nasa.gov/} resolution $N_{side}=512$). Since we are only interested in large angular scales we smooth all maps with appropriate filters to cut the power on small angular scales. These filters are low pass Gaussian filters
\be
W_{l}^{G} = N^{G}
\exp\left\{-\left(\frac{2l + 1}{2l_{s} + 1} \right)^{2}\right\}
\ee
that cut power on scales $(l \ge l_{s})$ and band pass filters of the
form
\be
W_{l}^{S} = 2 N^{S} \left[ 1 - J_{0}\left(\frac{2l + 1}
{2l_{t} + 1} \right)\right]\exp\left\{-\left(\frac{2l + 1}{2l_{s} + 1}
\right)^{2}\right\},
\ee
that keep the power on scales corresponding to $l_t < l < l_s$.$J_0$ is the  bessel function
and $N^{G}$ and $N^{S}$ are normalization constants chosen such that,
\be
\sum_{l}
\frac{(2l + 1)W_{l}}{2l(l + 1)} = 1
\ee
i.e., unit rms for unit flat band angular power spectrum $C_{l} = \frac{2\pi}
{l(l + 1)}$.
\begin{figure}[t]
\includegraphics[scale=0.3, angle=-90]{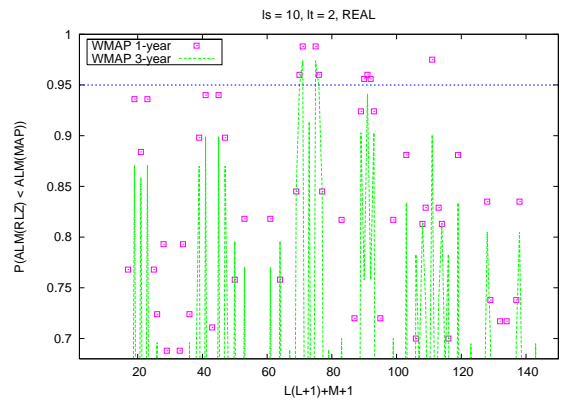}
    \caption{Fraction of $A_{\ell M}$'s  in 1000 simulations that are smaller than $A_{\ell M}$'s of ILC.
    Several deviations in ILC-1 (red square points) have been corrected in ILC-3 (green impulses).
     Only deviations above 68\% are shown. Almost all of $A_{\ell M}$'s of ILC-3 are smaller than those of ILC-1
     which means ILC-3 is more consistent with statistical isotropy. Only real parts are shown. Points are always in pairs because of
     the symmetry in reduced bipolar coefficients, $A_{\ell M}=(-1)^MA_{\ell-M}^*$. All maps are smoothed with a $W^S(10,2)$ filter. }
  \label{biggestALM}
\end{figure}

We compute the BipoSH coefficients, $\biposh{}{\ell M}{ll'}$, for 3-year ILC map (ILC-3) for several window functions using eqn.~(\ref{estimator}). We combine these coefficients using eqn.~(\ref{first}) to obtain $A_{\ell M}$.  
An interesting way of visualizing these coefficients is to make a map from them. Making a map from $A_{\ell M}$ is simply done similar to making a temperaure anisotropy map from a given set of spherical harmonic coefficients, $a_{lm}$;
\be
\Theta(\hat{n}) = \sum_{\ell=0}^{\infty}\sum_{M=-\ell}^{\ell} A_{\ell M} Y_{\ell M} (\hat{n}).
\ee
 The symmetry of reduced bipolar coefficients, eqn. (\ref{almsymmetry}), guarantees reality of $\Theta(\hat{n})$.
  The ``bipolar'' map based on bipolar coefficients of ILC-3 is shown on the top panel of  Fig. \ref{bipolarmap}. The map has small fluctuations except for a pair of hot and cold spots near the equator. To compare, we have also made a bipolar map of
 1-year ILC map (ILC-1) from bipolar coefficients of ILC-1 (middle panel of Fig. \ref{bipolarmap}). The difference map (Fig. \ref{bipolarmap} (bottom)) shows that differences between these two maps mostly arise from a band around the equator {\it in bipolar space}.
As it is seen in Fig. \ref{bipolarmap}, the bipolar map of ILC-3 has less fluctuations comparing to that of ILC-1. This is because almost all of $A_{\ell M}$'s of ILC-3 are smaller than those of ILC-1 ({\it i.e.} are closer to zero).  Reduced bipolar coefficients of the above maps are in Figure \ref{ALM_summed}, in which  $\ell$ and $M$ indices are combined to a single index $n=\ell(\ell+1)+M+1$ (only real part of $A_{\ell M}$ is plotted.). And the blue dotted lines define 1-$\sigma$ error bars derived from 1000 simulations of SI CMB anisotropy maps. As it can be seen many spikes presented in $A_{\ell M}$'s of ILC-1 have either disappeared or reduced in ILC-3 ({\it e.g.} those around $n=20, 40$ and a big spike at $n=111$). To get a quantitative description of differences between ILC-3 and ILC-1 we compare them against 1000 simulations of SI CMB anisotropy maps.  A simple $\chi^2$ comparison of $A_{\ell M}$ with simulations gives us a rough estimate of overall differences between the two ILC  maps:
ILC-3 has a smaller  $\chi^2$ than ILC-1. For a $W^S(10,2)$ filter,
the  reduced $\chi^2$ falls from $1.089 $ for ILC-1 to $0.9619$ for
ILC-3. Although  $\chi^2$ statistics is simple, it should be used
with caution because it is only valid if every   $A_{\ell M}$ is
independent has a Gaussian distribution function. In order to study
deviations of $A_{\ell M}$ from zero without worrying about the
Gaussianity of the $A_{\ell M}$, we look at the most deviant
(biggest) $A_{\ell M}$. We compare the biggest $A_{\ell M}$'s of ILC
to $A_{\ell M}$'s of 1000 simulations to find out what fraction of
simulations have $A_{\ell M}$'s smaller than those of ILC maps.
Figure \ref{biggestALM} shows the results. The horizontal axis is
$n=\ell(\ell+1)+M+1$ and the vertical axis is the fraction of
$A_{\ell M}$'s  in 1000 simulations that are smaller than $A_{\ell
M}$ of ILC. In this figure red squares represent the ILC-1 while
ILC-3 is represented by green lines. When a green line crosses a red
point,  $A_{\ell M}$'s of ILC-3 are greater than ILC-1, otherwise
red points above green spikes show smaller $A_{\ell M}$'s for ILC-3.
The results are interesting: several deviations in ILC-1 have been
corrected in ILC-3.  Specially on the largest scales, several
deviations beyound $95\%$ in ILC-1 have gone away in ILC-3 (red
points above the blue dotted line in Figure \ref{biggestALM} have
been replaced by significantly smaller values). However there are a
couple of exceptions that could be responsible for the hotter spot
in bipolar map of ILC-3.

\begin{figure}[t]
\includegraphics[scale=0.3, angle=-90]{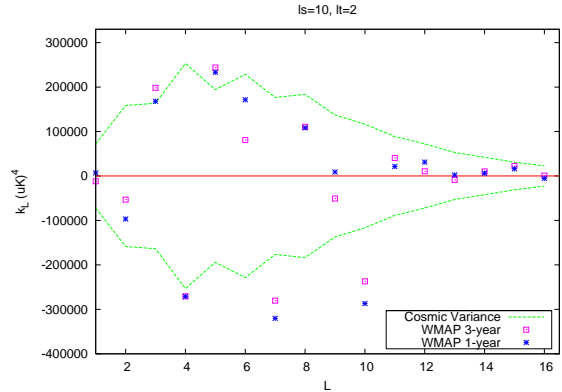}
    \caption{Bipolar power spectrum (BiPS) of the two ILC maps compared to average BiPS of 1000 simulations of statistically isotropic CMB maps. Both ILC maps are smoothed with a $W^S(10,2)$ window function which roughly retains multipoles in the range of $2\le l\le 15$. ILC-3 shows smaller BiPS than ILC-1, which means it is more consistent with statistical isotropy. Filtering the data with other functions show almost the same results.}
  \label{bips_fig}
\end{figure}

Combining the BipoSH coefficients to construct bipolar power
spectrum allows further examinations of ILC maps for departures from
SI. We compute the BiPS using eqn.~(\ref{estimatork}). It is worth
mentioning that BiPS in this paper has been computed in a slightly
different way than in our previous paper \cite{us_apj}. Here we
compute the BiPS using eqn.~(\ref{estimatork}) and we use the
derived $C_l$ from each map to estimate the  bias, ${\mathfrak
B}_\ell$, using eqn.~(\ref{klisobias})\footnote{For details of  bias
correction for BiPS see \cite{us_bigpaper} and \cite{us_bigpaper2}.}. The bias corrected
BiPS is then averaged over 1000 simulations and is compared to bias
corrected BiPS of ILC maps. BiPS results shown in Figure
\ref{bips_fig} agree with our results on $A_{\ell M}$. It can be
seen that ILC-3 has a smaller bipolar power spectrum than ILC-1 and
is more consistent with statistical isotropy. The same is true for
$D_{\ell}$ estimator defined by eqn. (\ref{dell}) which we defer to
the future publications. We should emphasize that these results are
only for large angular scales, $ l\le 25$, and not beyond that.

\section{Discussion and Conclusions}
\label{discussion} The null results of search for departure from
statistical isotropy has  implications for the observation and data
analysis techniques used to create the CMB anisotropy maps.
Observational artifacts such as  non-circular beam, inhomogeneous
noise correlation, residual striping patterns, and residuals from
foregrounds  are potential sources of SI breakdown.  Our null
results confirm that these artifacts do not significantly contribute
to large scale anisotropies of 3-year ILC map. We have also
quantified the differences between 1-year and 3-year ILC maps. It is
shown that 3-year ILC map is ``cleaner'' than 1-year ILC map at
$l\le 25$. This can be due to the gain model and improved ILC map
processing and foreground minimization. It has also been observed
that at large $l$  deviations from statistical isotropy occur which
we think is because of residuals from foregrounds. However we limit
ourselves to the low-$l$ limit because in addition to observational
artifacts, there are theoretical motivations for hunting for SI
violation on large scales of CMB anisotropy. Topologically compact
spaces~\cite{ell71, Lachieze-Rey:1995kj, lev02, linde} and
anisotropic cosmological models~\cite{ellis&maccallum,
collins&hawking1, collins&hawking2, doroshke, barrow, Jaffe:2006fh,
us_bianchi} are examples of this. Each of these models will cause
departures from statistical isotropy in CMB anisotropy maps. And a
null detection of departure from statistical isotropy at low $l$ in
the WMAP data can be used to put constraints on these models. Our
measure is sensitive to axial asymmetries in the two point
correlation of the temperature anisotropy \cite{us_prl}. And this is
even more significant now because the new measure of reduced bipolar
coefficients does retain directional information. Our analysis
doesn't show a significant detection of an ``axis of evil'' in the
WMAP data. We have redone our analysis on ILC map filtered with a
low-pass filter that only keeps $l=2,3,4$ to search for a preferred
direction at low multipoles. We have not been able to detect any
significant deviation from statistical isotropy using various
filters. We could not test the effect of alignment of low multipoles
on statistical isotropy because we had no theory or model to explain
them. Validity of statistical isotropy at large angular scales can
put tight constraints on anisotropic mechanisms that are candidates
of explaining the low quadrupole of the WMAP and COBE data. It is
worth noticing that our method can be extended to polarization maps
of CMB anisotropy. Analysis of statistical isotropy of full-sky
polarization maps of WMAP are currently under progress and will be
reported in a separate publication.

\section{Summary}
\label{summary} We examine statistical isotropy of large scale
anisotropies of the improved Internal Linear Combination (ILC) map,
based on three year WMAP data. In order to attribute a statistical
significance to our results, we use 1000 simulations of
statistically isotropic CMB maps. We have done our analysis using a
series of filters that span the low-$l$ multipoles. We only
explicitly present the results for one of them that roughly retains
power in the multipoles between 2 and 15. This reveals no
significant deviation from statistical isotropy on large angular
scales of 3-year ILC map. Comparing statistical isotropy of 3-year
ILC map and 1-year ILC map, we find a significant improvement in
3-year ILC map which can be due to the gain model and improved ILC
map processing and foreground minimization. We get consistent and
similar results from other filters.
\begin{acknowledgments}
AH wishes to thank Lyman Page and David Spergel for enlightening
discussions throughout this project. AH also thanks Soumen Basak for
his careful reading and comments on the manuscript and Joanna
Dunkley and Mike Nolta for useful discussions. Some of the results
in this paper have used the HEALPix package. We acknowledge the use
of the Legacy Archive for Microwave Background Data Analysis
(LAMBDA) \footnote{http://lambda.gsfc.nasa.gov/}. Support for LAMBDA
is provided by the NASA Office of Space Science. AH acknowledges
support from NASA grant LTSA03-0000-0090.
\end{acknowledgments}

\appendix
\section{Useful mathematical relations}
Bipolar spherical harmonics form an orthonormal basis of $S^2 \times
S^2$ and are defined as \be
\label{bipolars}
Y^{l_1l_2}_{\ell M}(\hat{n}_1,\, \hat{n}_2)\,=\, \sum_{m_1m_2}
{\mathcal C}_{l_1m_1l_2m_2}^{\ell M} Y_{l_1 m_1}(\hat{n}_1)Y_{l_2
m_2}(\hat{n}_2), \ee in which ${\mathcal C}_{l_1m_1l_2m_2}^{\ell M}$
are Clebsch-Gordan coefficients. Clebsch-Gordan coefficients are
non-zero only if triangularity relation holds, $\{l_1l_2\ell\}$, and
$M=m_1+m_2$. Where the $3j$-symbol $\{abc\}$ is defined by
\begin{displaymath}
\{abc\} = \left\{
\begin{array}{ll}
1 &\textrm{if $a+b+c$ is integer and $|a-b|\leq c \leq (a+b)$,}  \\
0 &\textrm{otherwise,}
\end{array} \right.
\end{displaymath}
Orthonormality of bipolar spherical harmonics \be \int
d\Omega_{\hat{n}_1}d\Omega_{\hat{n}_2} \, Y^{l_1l_2}_{\ell
M}(\hat{n}_1,\, \hat{n}_2) Y^{*l'_1l'_2}_{\ell' M'}(\hat{n}_1,\,
\hat{n}_2) = \delta_{l_1l'_1} \delta_{l_2l'_2}\delta_{\ell
\ell'}\delta_{MM'} \label{A2} \ee

\end{document}